# Highly Accurate Measurement of the Electron Orbital Magnetic Moment

A. M. Awobode

Department of Physics, University of Massachusetts, Boston, MA 02125, USA.

**Abstract**

We propose to accurately determine the orbital magnetic moment of the electron by measuring, in a Magneto-Optical or Ion trap, the ratio of the Lande g-factors in two atomic states. From the measurement of $(g_{J1}/g_{J2})$, the quantity $A = (\alpha\delta_S - \beta\delta_L)$ can be extracted, if the states are LS coupled. The indirectly measured quantity is a linear combination of the $\delta_S$ and $\delta_L$ which are, respectively, the corrections to the spin and orbital g-factors where $\alpha$, $\beta$ are constants. Given that highly accurate values of $\delta$s are currently available, accurate values of $\delta_L$ may also be determined. At present, the correction $\delta_L = (-1.8 \pm 0.4) \times 10^{-4}$ has been determined by using earlier measurements of the ratio of the g-factors, made on the Indium $^2P_{1/2}$ and $^2P_{3/2}$ states.

E-mail: awobode@gmail.com

**Introduction: Atomic Beam Magnetic Resonance Experiments**

The classic experiments of Kusch et al [3-6], using the Atomic Beam Zeeman Resonance technique measured the quantity $(\alpha\delta_S - \beta\delta_L)$ to about 7 parts in $10^5$. The results of these experiments are currently of interest, because of the unprecedented, recent measurement of the $\delta$s by Gabrielse et al [1]. By studying the cyclotron motion of an electron confined in a Penning trap, Gabrielse et al measured the correction to the spin g-factor to 1 part in $10^{12}$, which is a great improvement on the earlier measurement by Dehmelt et al [2], who measured the same quantity to 1 part in $10^9$. One of the reasons for the renewed interest in the measurement of $(\alpha\delta s - \beta\delta_L)$ is the possibility of determining $\delta_L$, given that $\delta$s is known to such great accuracy. For example, combining the work of Mann and Kusch [3] with the recent result of Gabrielse et al indicates that $\delta_L = (-1.8 \pm 0.4) \times 10^{-4}$. This is significantly more accurate than $(-0.6 \pm 0.3) \times 10^{-4}$, the first estimate of $\delta_L$ [7], based on the earlier work of Kusch and Foley [4]. While these clearly show that $\delta_L \neq 0$, more accurate values may be determined using laser-cooled atoms in Magneto-Optical traps.

**Measurement of the Electron Gyromagnetic Factors**

Assuming Russell-Saunders (LS) coupling, the ratio of the g-factors,

$$g_{JI}/g_{J2} = (a_{L1}g_L + a_{S1}g_S)/(a_{L2}g_L + a_{S2}g_S)$$

may be expressed as [5],

$$g_{J1}/g_{J2} = [(2a_{S1} + a_{L1})/(2a_{S2} + a_{L2})] + 2[(a_{S1}*a_{L1} - a_{L2}*a_{S2})/(2a_{S2} + a_{L2})^2](\delta_L - \delta_S)$$

where it is supposed that the orbital and spin g-factors are corrected as follows:

$$g_S = 2(1 + \delta_S) \quad \text{and} \quad g_L = 1 + \delta_L$$

Thus, if all the constants in the above equations are known, then the quantity $(\delta_L - \delta_S)$ or some other linear combination of the $\delta_S$ and $\delta_L$ may be determined, where the constants

$$a_L = [J(J+1) + L(L+1) - S(S+1)]/J(J+1) \text{ and } a_S = [J(J+1) - L(L+1) + S(S+1)]/J(J+1)$$

are based on LS coupling.

Using atoms of Na, Ga and In, Kusch et al measured the frequency of the transitions that showed the greatest magnetic sensitivities. The frequencies were determined to about 1 part in 20,000. The magnetic field employed was typically about 400 gauss. In principle, it is possible from the observed frequencies of the line spectra of atoms in two states to calculate directly the ratio $g_{J1}/g_{J2}$. However, such a procedure is laborious and instead, Kusch et al calculated the quantity $H´ = 1.3998H$ for each observed line, and determined from

$$g_{J1}/g_{J2} = (g^0_{J1}/g^0_{J2})[1 + \Delta H´/H´]$$

the ratio of the Lande g-factors. Here, $\Delta H´ = H´_1 - H´_2$ and $g^0_{J1}$ and $g^0_{J2}$ are the assumed (LS-coupled) values of $g_J$ for two different states. The results of the measurements are described in the table below.

**Table 1.** Measurement of the ratio of Lande g-factors & the determination of $\delta_S$ assuming $\delta_L = 0$

| Reference | Atomic states and gJ | Measured ratios of gJ | $\delta_S$ ($\delta_L = 0$ is assumed) |
|---|---|---|---|
| Kusch and Foley(1947) | $g_{J1}(^2P_{3/2}Ga)/g_{J2}(^2P_{1/2}Ga)$ | $2(1.00172 \pm 0.00006)$ | $0.00114 \pm 0.00004$ |
| Foley and Kusch (1947) | $g_{J1}(^2S_{1/2}Na)/g_{J2}(^2P_{1/2}Ga)$ | $3(1.00242 \pm 0.00006)$ | $0.00121 \pm 0.00003$ |
| Kusch and Foley (1948) | $g_{J1}(^2S_{1/2}Na)/g_{J2}(^2P_{1/2}In)$ | $3(1.00243 \pm 0.00010)$ | $0.00121 \pm 0.00005$ |
| Mann and Kusch (1950) | $g_{J1}(^2P_{3/2}In)/g_{J2}(^2P_{1/2}In)$ | $2(1.00200 \pm 0.00006)$ | $0.00133 \pm 0.00004$ |

From the quantities in **Table 1**, the earlier values of $\delta_S$ were calculated on the assumption that $\delta_L = 0$ exactly. It is however possible to make a better estimate of the $\delta_L$ given that high-precision value of $\delta_S$ is now available. The value given by Gabrielse et al is:

gs/2 = 1.001 159 652 180 73 (28) [0.28 ppt]

which, to three orders of magnitude, is better than that of Dehmelt et al, given as:

gs/2 = 1.001 159 652

Putting together the results in **Table 1** and the values of δs given by Gabrielse et al or Dehmelt et al, we find $\delta_L$.

**Table 2.** Determination of $\delta_L$ from the measured values of $A = \alpha\delta s - \beta\delta_L$ and δs.

| $A = \alpha\delta s - \beta \delta l$ | $\delta_S$ | $\delta_L$ |
|---|---|---|
| 1. $\delta s - 2\delta l = 0.00229 \pm 0.00008$ | $\delta_S = 0.00231$ | $(-0.1 \pm 0.4) \times 10^{-4}$ |
| 2. $\delta s - 2\delta l = 0.00244 \pm 0.00006$ | $\delta_S = 0.00231$ | $(-0.6 \pm 0.3) \times 10^{-4}$ |
| 3. $\delta s - 2\delta l = 0.00243 \pm 0.00006$ | $\delta_S = 0.00231$ | $(-0.6 \pm 0.3) \times 10^{-4}$ |
| 4. $3\delta s - 3\delta l = 0.00400 \pm 0.00012$ | $\delta s = 0.00115$ | $(-1.8 \pm 0.4) \times 10^{-4}$ |

Thus, $\delta_L$ takes the values $-1.8 \times 10^{-4}$, $-0.6 \times 10^{-4}$ and $-0.1 \times 10^{-4}$. It is noteworthy that the order of magnitude and the sign are consistent. All experimental data in the measurements of Kusch et al are measurement of spectral lines; no knowledge of the magnetic field in which the transitions occur is required, although it is necessary that the transitions resulting in lines whose frequencies are to be compared occur in the same magnetic field. In practice however, the magnetic field is not entirely constant and the drift depends on a number of factors; nevertheless, a correction for the drift can be made. Also, the uncertainty in frequency measurement imposed by the frequency meter and the line widths are statistical in character, hence precision may be improved by judicious repetition of observations.

There are perturbations of the electronic states involved in the experiments which in principle could bring about deviations of the atomic values from that given by the Russell-Saunders coupling formula. However, they are negligibly small ($< 1 \times 10^{-5}$). Prominent among these is the electrostatic interaction, which mixes states of the same values of total L and S and therefore does not change the g value in any approximation. Also, the effect on the $g_J$ values of the magnetic interactions and spin-orbit interactions are negligible; the effect of the magnetic interactions may however be larger In than in Ga. The effect of configuration interactions on the $g_J$ values of the ground states of alkali metals was investigated by Phillips who concluded that these effects are negligible. Relativistic effects on the $g_J$ values of a Dirac electron have also been considered, and for the alkali metals, the reduction is about 1 part in $10^5$, and for In and Ga, the reduction is about 4 parts in $10^5$. The magnitude of the orientation dependence of the magnetic susceptibility is also negligible (~1 part in $10^5$) at the field employed in the experiments. Incipient electronic Paschen-Back effect produces a perturbation in the line frequencies of the $^2P_{3/2}$ (Ga) of about one part in $10^5$. Also, the interaction of the valence electron spin with the diamagnetically induced moment of the core electrons reduces the $g_J$ values by less than 1 part in $10^5$. Despite the possible presence of perturbations, Kusch and Foley concluded that the observed ratios $g_{J1}/g_{J2}$ (Na/Ga) and $g_{J1}/g_{J2}$ (Na/In) cannot be due to perturbations, because all the states cannot be subject to perturbations such that the values of the ratios are almost the same: *"The agreement between the values of ($\delta_S - 2\delta_L$) obtained by the two experiments makes it unlikely that one can account for the effect by perturbation of the states. The effect of configuration interaction on the $g_J$ value of Na is presumably negligible. To explain our observed effect without modification of the conventional values of $g_S$ or $g_L$ introduces the rather unlikely requirement that both states of Ga (and In) be perturbed, and by amounts just great enough to give the agreement above"* [5].

The application of all corrections to the ratio of the g-factors affects the calculated values of $\delta_S$ by about 1%. Thus the best value of $g_S$ from the experiments of Kusch and Foley is $g_S = 2(1.00119 \pm 0.00005)$. However, on the assumption that $\delta_L = 0$ and $\delta_S$ is calculated, Kusch and Foley remarked that "The discrepancies between the individual values of $\delta_S$ indicate the existence of small residual systematic effects…" [5]. Considering the highly accurate values of $\delta_S$ currently available, the "residual systematic effect" may, in fact, be due to corrections to the orbital g-factor, for the states as measured.

**Discussion and Conclusion**

Greatly improved measurements of the ratio ($g_{J1}/g_{J2}$) from which may be derived the quantity ($\delta_S - 2\delta_l$) can be obtained by using confined, cold atoms. Traps for charged and neutral particles have been used for the precise determination of fundamental particle properties. The extended observation time for stable particles under well controlled conditions and the low energy to which the stored particles can be cooled reduces the uncertainties to subparts per billion. For example, to determine $\delta_S$ in bound systems, Werth et al were able to measure, using microwave spectroscopy, the $g_J$ values in the $l = 0$ states of highly charged ions [8] to an accuracy of about $1.1 \times 10^{-9}$; however, experiments have reached a level of precision where effects depending on the size of the trap limit further progress. Although, measurements could (in principle) be made on $l \neq 0$ states of ions, laser-cooled neutral atoms in magneto-Optical traps offer the possibility of accurate measurement on atomic systems, which are modernized versions of the Kusch-Foley experiments. To date, over 30 atoms (including Na, Rb and In)* have been laser cooled and trapped and recently the list have included Ho, Dy and Er [10]. Also, high resolution microwave and optical spectroscopy measurements can be performed simultaneously on ions and atoms. Vogel et al [9] have recently described a new technique called the laser-microwave double resonance with which $g_J$ values of bound systems may be determined to an accuracy in the range of $1 \times 10^{-9}$. An accuracy of about 1 part in $10^7$ will be sufficient to establish the presence or otherwise of a non-vanishing correction to the orbital g-factor. Deleterious effects of perturbations on the measured $g_J$ of atomic states may be ascertained by the observations of abnormality in the series, interval ratios and intensities, which are also affected by perturbations.

Modern versions of the experiments by Kusch et al are required in order to determine to an accuracy of $1 \times 10^{-7}$ or better, the correction to the electron orbital g-factor. Following the theory of experiments outlined by Kusch and Foley, it is possible to obtain from a measurement of the ratio $(g_{J1}/g_{J2})$, the quantity $(\alpha\delta_L - \beta\delta s)$ from which $\delta_L$, the correction to $g_L$ can be determined, given that $\delta s$ is known to better than 5 parts in $10^{12}$. It is necessary to select states (in atoms or ions) which conform to Russell Saunders (LS) coupling in agreement with the theory of the experiment. Such states should also be free, as much as possible, from perturbations. Using high resolution microwave spectroscopy, $\delta_L$ can be measured to a few parts in a billion.